\documentclass[twocolumn,amsmath,amssymbprl]{revtex4}

\usepackage{graphicx}
\usepackage{dcolumn}
\usepackage{bm}

\begin{document}


\title{The origin of the positron excess in cosmic rays}

\author{Pasquale Blasi}
\affiliation{INAF/Osservatorio Astrofisico di Arcetri, 
Largo E. Fermi, 5 50125 Firenze (Italy)}

\date{\today}

\begin{abstract}
We show that the positron excess measured by the PAMELA experiment in the region between 10 and 100 GeV may well be a natural consequence of the standard scenario for the origin of Galactic cosmic rays. The 'excess' arises because of positrons created as secondary products of hadronic interactions inside the sources, but the crucial physical ingredient which leads to a natural explanation of the positron flux is the fact that the secondary production takes place in the same region where cosmic rays are being accelerated. Therefore secondary positrons (and electrons) participate in the acceleration process and turn out to have a very flat spectrum, which is responsible, after propagation in the Galaxy, for the observed positron 'excess'. This effect cannot be avoided though its strength depends on the values of the environmental parameters during the late stages of evolution of supernova remnants. 
\end{abstract}
\pacs{Valid PACS appear here}
\maketitle
The PAMELA satellite began its three-year mission in June of 2006, and among its goals there was that of  measuring the spectra of cosmic ray positrons up to 270 GeV and electrons up to 2 TeV, each with unprecedented precision \cite{pamsite}. Recent results \cite{adriani} show that the ratio of positrons to electrons plus positrons (the so-called positron fraction) in the cosmic ray spectrum appears to rise with energy, at least up to $\sim 100$ GeV, as already found by previous experiments, including HEAT \cite{heat} and AMS-01 \cite{ams01}, although with smaller statistical significance. The clear discrepancy between the observed positron fraction and the predictions of the standard model for the origin and propagation of cosmic rays in the Galaxy, led to many possible explanations, ranging from the annihilation of non-baryonic dark matter \cite{dark} to the possibility that new astrophysical sources, especially pulsars \cite{pulsars}, could provide the additional positron flux. It is worth recalling that both lines of thought lead to the ``correct'' spectral slope rather naturally, but they are very different in terms of providing the correct normalization of the positron fluxes. The dark matter interpretation typically requires large, and somewhat artificial, annihilation rates. Such large rates could, in principle, result from dark matter possessing an annihilation cross section in excess of the value predicted for a simple s-wave thermal relic ($\sigma v \approx 3 \times 10^{-26} cm^{3}/sec$), for example due to the Sommerfeld effect \cite{som1}, or from non thermal WIMPs \cite{nonthermal}. In the case of pulsars on the other hand, the energetic requirements appear to be all but extreme, although an efficiency factor needs to be introduced by hand \cite{pulsars}, and the mechanisms for escape of the pairs from the pulsar environment are basically unknown.

At this point it is useful to recall what are the predictions for the positron flux of the so-called 'standard model' of the cosmic ray propagation in the Galaxy: cosmic rays are assumed to be accelerated in astrophysical sources, such as supernova remnants (SNRs), with a source spectrum $N_{CR}(E)\propto E^{-\gamma}$ (throughout the paper we adopt the convention that for any particle distribution function $f$ one has $4\pi p^{2} f(p)dp = f(E) dE$). The power law behaviour, in the case of SNRs, naturally arises from diffusive shock acceleration \cite{bland}. The spectrum of cosmic rays observed at Earth is $n_{CR}(E)\propto N_{CR}(E)\tau_{esc}(E)$, where $\tau_{esc}(E)$ is the escape time from the Galaxy. This time scale is inferred from the ratio of Boron to Carbon (B/C) fluxes in cosmic rays and is typically found to scale as $\tau_{esc}(E)\propto E^{-\delta}$, with $\delta\approx 0.3-0.6$. In this standard model, positrons only arise as secondary products of cosmic ray interactions in the Galaxy. Positrons are therefore injected at a rate $Q_{+}(E)\propto n_{CR}(E) n_{H} c \sigma$ (where $\sigma$ is the cross section for the relevant process), while the equilibrium spectrum is  $n_{+}(E)\propto Q_{+}(E)\tau_{e}(E)\propto E^{-\gamma-\delta}\tau_{e}(E)$, where $\tau_{e}(E) \approx Min[\tau_{esc}(E),\tau_{loss}(E)]$, and $\tau_{loss}$ is the loss time scale during propagation in the Galaxy (this scaling may however be affected by the spatial distribution of the sources with respect to the diffusion region). In the range of energies we are interested in, in most cases energy losses are dominant over escape, therefore $\tau_{e}(E)\approx \tau_{loss}(E)\propto 1/E$ (losses are dominated by inverse Compton scattering (ICS) and synchrotron emission). It follows that  the spectrum of positrons observed at Earth is $n_{+}(E)\propto E^{-\gamma-\delta-1}$. The injection spectrum of electrons in SNRs is usually parametrized as $N_{e}(E) = K_{ep} N_{CR}(E)$, so that following the same reasoning the equilibrium spectrum of primary cosmic ray electrons is $n_{-}(E)\propto E^{-\gamma-1}$. It is straightforward to see that the ratio $n_{+}/n_{-}\propto E^{-\delta}$, namely it decreases with energy, at odds with PAMELA findings. These simple arguments are nicely summarized in \cite{serpico}. 
In this paper we show that the standard model as summarized above lacks one important phenomenon which is intrinsic in the acceleration process. We also show how taking into account this phenomenon a positron excess arises in the standard picture of acceleration of cosmic rays in Galactic sources.
The spectrum and spatial distribution of cosmic rays in the accelerator are obtained by solving the transport equation \cite{bell78,bland} and one finds the well known result that the solution is 
\begin{equation}
f_{CR} (x,p) = K \left( \frac{p}{p_{0}} \right)^{-\gamma} F(x,p),
\end{equation}
where $p$ is the particle momentum, $F(x,p)=1$ downstream ($x>0$) and $F(x,p)=\exp\left(ux/D(p)\right)$ upstream ($x<0$). The predicted cosmic ray spectrum at the shock surface ($x=0$) is a power law in momentum and the slope is $\gamma = 3u_{1}/(u_{1}-u_{2})=3r/(r-1)$, where $u_{1,2}$ are the fluid velocities upstream and downstream and $r=u_{1}/u_{2}$ is the compression factor. For a strong shock $r\to 4$ and $\gamma\to 4$. For relativistic particles ($p\sim E$) the spectrum in energy is $N_{CR}(E)=4\pi p^{2} f_{CR}(p) u_{2}\tau_{SN}\sim E^{-\gamma+2}$, and for a strong shock $N_{CR}(E)\sim E^{-2}$, the well known result. Here $\tau_{SN}$ is a typical age of a SNR.

Cosmic rays accelerated at the shock produce secondary $e^{-}+e^{-}$ inside the source through hadronic interactions with production and decay of charged pions, while the decay of neutral pions leads to production of gamma rays. It is crucial to realize that this process occurs in the same spatial region around the shock in which cosmic rays are being accelerated. Therefore it is unavoidable that secondary $e^{-}+e^{-}$ take part in the same acceleration process. Notice that these particles are already suprathermal, therefore we do not need to worry about their injection. The production rate at a position $x$ around the shock is 
\begin{equation}
Q_{\pm}(x,E) = \int dE' N_{CR}(E',x) \frac{d\sigma(E',E)}{dE'} n_{gas}(x) c ,
\end{equation}
where $c$ is the speed of light, $n_{gas}$ is the gas density for $pp$ scattering in the shock region ($n_{gas,2}=r n_{gas,1}$) and $\frac{d\sigma_{\pm}(E',E)}{dE'}$ is the differential cross section for a proton of energy $E'$ to produce an $e^{+}$ or an $e^{-}$ of energy $E$. Here we calculate these cross sections following the approach of \cite{kamae}.

The calculations are carried out neglecting the non-linear effects which lead to the appearance of a precursor upstream of the shock (see \cite{maldru} for a review). Moreover we do not comment here on some important aspects of the escape of particles from the remnant which are as relevant here as for ordinary cosmic rays, but imply a difficult integration over time of the acceleration history.

The most important point of this paper is the calculation of the equilibrium spectrum of $e^{-}+e^{-}$ produced in the acceleration region. Their behavior is described by the transport equation which automatically takes into account the presence of the shock, the advection with the fluid and diffusion:
\begin{equation}
u \frac{\partial f_{\pm}}{\partial x} = D(p) \frac{\partial^{2} f_{\pm}}{\partial x^{2}} + 
\frac{1}{3}\frac{du}{dx}p\frac{\partial f_{\pm}}{\partial p} + Q_{\pm}(x,p),
\label{eq:trans}
\end{equation}
where $f_{\pm}(x,p)$ is the equilibrium distribution function of electrons (-) and positrons (+) (number of particles per unit volume per unit energy). In the equation above we neglect the effect of energy losses of electrons because here we will be restricting ourselves to situations in which the secondary pairs have energy below that for which synchrotron and ICS losses are important. We will check {\it a posteriori} that this is the case.

Eq. (\ref{eq:trans}) has to be solved with the following boundary conditions: $f_{\pm}(x,p)$ and $\partial f_{\pm}/\partial x$ vanish at upstream infinity ($x\to -\infty$) and $\partial f_{\pm}/\partial x$ remains finite at downstream infinity ($x\to +\infty$). The boundary condition at the shock surface is obtained in a straightforward way by simply integrating Eq. (\ref{eq:trans}) in a narrow region around the shock. With these conditions, one can easily show that the solution to Eq. (\ref{eq:trans}) 
is in the form 
\begin{equation}
f_{\pm}(x,p)=f_{\pm,0}(p)+\frac{Q_{2}}{u_{2}}~x,
\label{eq:solution}
\end{equation}
where $f_{\pm,0}(p)=f_{\pm}(x=0,p)$ and is the solution of the following equation:
\begin{equation}
p \frac{\partial f_{\pm,0}}{\partial p} = -\gamma f_{\pm,0} + \gamma (\frac{1}{\xi}+r^{2}) \frac{D_{1}(p)}{u_{1}^{2}}  Q_{1}(p),
\label{eq:trans1}
\end{equation}
where $D_{1}(p)\propto p^{\alpha}$ is the diffusion coefficient upstream of the shock, and $Q_{1}(p) = Q_{\pm}(x=0^{-},p)$ is the rate of injection of pairs immediately upstream of the shock. The factor $\xi\sim 0.05$ represents the mean fraction of the energy of an accelerated proton carried away by a positron or electron in each scattering. 

The solution of this equation is promptly found to be:
\begin{equation}
f_{\pm,0} (p) = \gamma (\frac{1}{\xi}+r^{2}) \int_{0}^{p} \frac{dp'}{p'} \left( \frac{p'}{p} \right)^{\gamma} \frac{D_{1}(p')}{u_{1}^{2}} Q_{1}(p').
\label{eq:sol}
\end{equation}
The physical meaning of Eq. (\ref{eq:solution}) is that secondary particles which are produced within a distance $\sim D(p)/u$ from the shock (on both sides) participate in the acceleration process (first term in Eq. (\ref{eq:solution})), while the downstream secondary particles produced farther away are simply advected and their density increases with $x$ (second term in Eq. (\ref{eq:solution})) up to a maximum distance $\sim u_{2}\tau_{SN}$, where $\tau_{SN}$ is the age of the SNR. The contribution of SNRs to the flux of secondary $e^{-}$ and $e^{+}$ is the integral of Eq. (\ref{eq:solution}) over $0\leq x\leq u_{2}\tau_{SN}$.

One should keep in mind that roughly $Q_{1}(p)\propto f_{CR}(p)\sim p^{-\gamma}$ (at least at $p>1GeV/c$), therefore Eq. (\ref{eq:sol}) leads to $f_{\pm,0}\sim p^{-\gamma+\alpha}$, where $\alpha>0$ is the slope of the diffusion coefficient as a function of energy: the equilibrium spectrum of the pairs that take part in the acceleration is flatter than the injection spectrum of secondary pairs. 
This is crucial to understand that the flux of secondaries in the sources cannot be simply rescaled with grammage from the interstellar medium cosmic ray interactions because of the strong dependence of  positron production on the spectrum of primaries. 

At this point we can calculate the equilibrium spectra of primary electrons and secondary $e^{-}+e^{-}$ in the Galaxy after propagation. We model the Galaxy as a cylinder with half-height $H=3$ kpc. We adopt a diffusion coefficient in the Galaxy $D_{gal}(E)=D_{0}(E/GeV)^{\delta}$, with $\delta=0.6$ and $D_{0}$ chosen so to have an escape time at 10 GeV of $\tau_{esc}(10 GeV)\approx 20$ Myr: $D_{0} \approx 10^{28}cm^{2} s^{-1}$. Energy losses of electrons and positrons are due to ICS and synchrotron emission and we adopt a typical total rate of energy losses $b(E)=\frac{dE}{dt}\approx 2\times 10^{-16}E_{GeV}^{2}~GeV/s$. The simplified picture that arises is leaky-box-like with all limitations that it implies.

If ${\cal R_{SN}}$ is the rate of supernovae occurring in our Galaxy per unit volume, the equilibrium distribution of cosmic rays in the Galaxy is
\begin{equation}
n_{CR}(E) = N_{CR}{\cal R_{SN}}\tau_{esc}(E). 
\end{equation}
The equilibrium spectrum of secondary $e^{-}+e^{-}$ produced by cosmic ray interactions in the Galaxy is determined by a balance between injection, losses and escape from the Galaxy. For the diffusion coefficient $D(E)\approx 10^{28}E_{GeV}^{0.6}cm^{2}s^{-1}$ the loss time is shorter than the escape time at all energies above $\sim 10$ GeV, namely at all energies of interest for us. In this case the equilibrium spectrum of the diffuse secondary pairs can easily be written as 
\begin{equation}
n_{\pm}(E) = \frac{K_{N} n_{H} c}{b(E)} \int_{E}^{E_{max}} dE'' \int dE' n_{CR}(E') \frac{d\sigma_{\pm}(E',E'')}{dE'},
\end{equation}
where $n_{H}$ is the gas density averaged over the volume of the Galaxy (including disc and halo) and a coefficient $K_{N}\sim 1.2-1.8$ is introduced to account for the interaction of nuclei other than hydrogen. Following \cite{KN} we use $K_{N}=1.8$. Clearly, the choice of a different diffusion coefficient in the Galaxy may lead to the need for a more detailed solution, taking into account the interplay between escape and losses. Moreover if a non-leaky box model is used, a slightly different slope of the equilibrium spectra is obtained, though the positron fraction remains unaffected.

Similarly, for the secondary pairs produced inside the sources, one has:
\begin{equation}
n_{\pm}^{s}(E) = K_{N} {\cal R_{SN}}\frac{1}{b(E)} \int_{E}^{E_{max}} dE' N_{\pm}^{s}(E'),
\end{equation}
where $N_{\pm}^{s}(E)dE=4\pi p^{2} \left[f_{\pm,0}+(1/2)Q_{2}\tau_{SN}\right] u_{2}\tau_{SN}dp$ is the distribution function of pairs at the sources in energy space instead of momentum space (we integrated Eq. (\ref{eq:solution}) over the downstream volume, exactly as for CRs). 

Finally, for the spectrum of primary electrons in the sources we adopt the standard procedure of assuming that $N_{e}(E)=K_{ep} N_{CR}(E)$, where $K_{ep}\approx 7\times 10^{-3}$. The equilibrium spectrum of primary electrons is then:
\begin{equation}
n_{e}(E) = K_{ep}{\cal R_{SN}} \frac{1}{b(E)} \int_{E}^{E_{max}} dE' N_{CR}(E').
\end{equation}
Before illustrating the results of our calculations we discuss briefly the choice of diffusion coefficient in the accelerator, which is not the same as in the Galaxy, because of the generation (and damping) of turbulence in the shock region, either due to the same accelerated particles \cite{bell78} or due to fluid instabilities. Here we carry out the calculations for a Bohm-like diffusion coefficient, which we write as:
\begin{equation}
D_{B}(E)=K_{B} \frac{1}{3} r_{L}(E) c=3.3\times 10^{22}K_{B}B_\mu^{-1}E_{GeV} ~ cm^{2}s^{-1}.
\end{equation}
Here $B_\mu$ is the local ordered magnetic field in units of $\mu G$ and the coefficient $K_{B}\simeq (B/\delta B)^{2}$ allows to consider faster diffusion ($K_{B}>1$), which is common when magnetic field amplification is not as efficient. 

\begin{figure}
\begin{center}
{\includegraphics[angle=0,width=1.\linewidth]{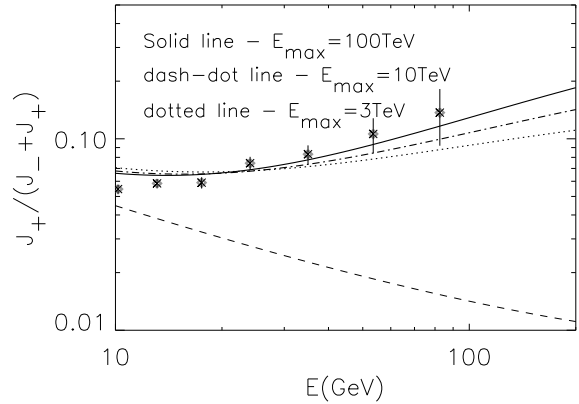}}
\caption{Positron fraction as a function of energy. The data points are the results of the PAMELA measurement.}
\label{fig:ratio}
\end{center}
\end{figure}

These are all the ingredients needed for the calculation of the positron and electron fluxes at Earth. The positron fraction, defined as the ratio of the total flux of positrons to the total flux of $e^{-}+e^{+}$, is plotted in Fig. \ref{fig:ratio}. The data points are the results of the PAMELA measurement. The error bar on energy is of the order of half the distance between two consecutive data points. 
The solid line refers to the case of maximum energy of the accelerated particles (and therefore also of the secondary particles after reacceleration) $E_{max}=100$ TeV, while the dash-dotted and dotted lines refer respectively to $E_{max}=10$ TeV and $E_{max}=3$ TeV. The dashed curve represents the standard contribution to the positron fraction from secondary diffuse pairs. We adopt a reference age $\tau_{SN}\approx 10^{4}$ years for a SNR. The three curves refer to $\{K_{B},n_{gas,1},B_{\mu},u_{8}\}=\{20,1.3,1,0.5\}$ for $E_{max}=100$ TeV, $\{20,2,1,0.5\}$ for $E_{max}=10$ TeV, and $\{20,3,1,0.5\}$ for $E_{max}=3$ TeV ($n_{gas,1}$ is the gas density close to the SNR in units of $1cm^{-3}$ and $u_{8}=u_{1}/10^{8}cm/s$). One can see that these values are appropriate for old supernova remnants, which however are also expected to be the ones that contribute the most to the cosmic ray flux below the knee. Unfortunately during such phase the maximum energy of accelerated particles decreases in time in a way which is very uncertain: slowly in the case of no damping and rather fast if effective magnetic field amplification and damping are present. This is the reason why in Fig. \ref{fig:ratio} we considered the three values of $E_{max}$. A solid evaluation of this effect can only be achieved by carrying out a fully time dependent calculation (Caprioli and Blasi, in preparation). A prediction of this scenario is that the positron fraction grows and eventually levels out at $\sim 40-50\%$. 
\begin{figure}
\begin{center}
{\includegraphics[angle=0,width=0.9\linewidth]{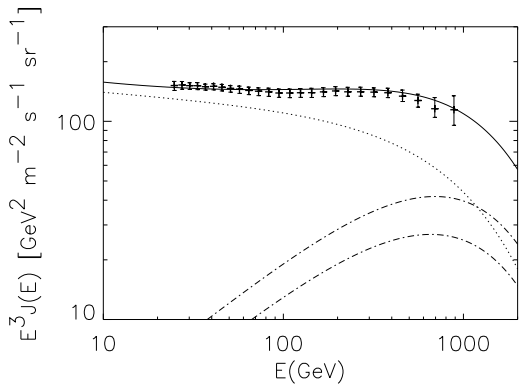}}
\caption{Fluxes of $e^{-}$ and $e^{+}$ at Earth for $E_{max}=100$ TeV. The dotted line refers to primary electrons, the dashed lines are the fluxes of positrons (upper curve) and electrons (lower curve) from interactions of cosmic rays in the Galaxy. The dot-dashed lines are the fluxes of positrons (upper curve) and electrons (lower curve) from production in the sources. The thick solid line is the total flux. The data points are from Fermi/LAT \cite{fermi}.}
\label{fig:flux}
\end{center}
\end{figure}
The fluxes of electrons and positrons are plotted in Fig. \ref{fig:flux} for the case $E_{max}=100$ TeV. We assumed that the closest source of cosmic rays is located at a distance of order $\sim 1-2$ kpc, so to introduce a high energy cutoff at $\sim 1$ TeV, namely when the propagation time from the closest source exceeds the loss time (this is a strong function of the distance to the closest source). A cutoff may also be produced by the acceleration process in the sources. Recent observations by ATIC \cite{atic}, HESS \cite{hess_ele} and Fermi/LAT seem to confirm the presence of the high energy cutoff in the diffuse electron spectrum. The dotted line refers to primary electrons, the dashed lines are the fluxes of positrons (upper curve) and electrons (lower curve) from interactions of cosmic rays in the Galaxy. The dot-dashed lines are the fluxes of positrons (upper curve) and electrons (lower curve) from production in the sources. The data points are from Fermi/LAT \cite{fermi}. A few remarks are in order: 1) at energies above $\sim 20$ GeV the main contribution to the positron flux is strikingly the one of secondary pairs in the sources. 2) The flat spectrum of the secondary pairs, energized inside the acceleration region, makes them provide up to $50\%$ of the total flux of electrons (and positrons) at Earth at high energy. Their contribution is sufficient to flatten the total $e^{+}+e^{-}$ spectrum, which is the quantity effectively measured by ATIC and Fermi/LAT. 3) No sharp feature appears in the total spectrum, at odds with ATIC data, which are however not confirmed by Fermi/LAT results. 

We also stress that a contribution to this positron flux might come from a fraction of SNRs located in proximity of dense molecular clouds, where the target for $pp$ collisions may be enhanced and the importance of the mechanism made more evident. These SNRs might also have sufficiently high surface brightness (despite the old age) to be detected by Fermi/LAT, provided the target density for $pp$ interactions is large enough. 

We can conclude that the positron excess can be a consequence of acceleration of cosmic rays in SNRs or other sources. The scenario discussed here has numerous implications: first, a flux of antiprotons is predicted to be produced, which is compatible with present data \cite{antip}; second, a similar effect, though more model dependent, appears in the spectra of secondary nuclei which are now being calculated. As far as individual sources are concerned, one has to keep in mind that no effect is expected from young, bright SNRs. In old remnants a possible signature might appear in radio spectra but may be easily confused with other effects. In any case supernovae of type Ia are more likely to contribute to this effect because they typically explode in denser media. 

\begin{acknowledgments}
The author is grateful to R. Aloisio, E. Amato, D. Caprioli, G. Morlino and P. Serpico for endless discussions and exciting collaboration. 
\end{acknowledgments}


\begin{thebibliography}{99}
\bibitem{pamsite}
http://pamela.roma2.infn.it/index.php.

\bibitem{adriani}
O. Adriani {\it et al.}, Preprint arXiv:0810.4995.

\bibitem{heat}
S. W. Barwick {\it et al.} [HEAT Collaboration], Astrophys. J. {\bf 482}, L191 (1997); S. Coutu et al.
[HEAT-pbar Collaboration], in Proceedings of 27th ICRC (2001).

\bibitem{ams01}
M. Aguilar {\it et al.} [AMS-01 Collaboration], Phys. Lett. {\bf B646}, 145 (2007).

\bibitem{dark}
I. Cholis {\it et al.}, Preprint arXiv:0809.1683.

\bibitem{pulsars}
D. Hooper, P. Blasi and P.D. Serpico, JCAP {\bf 1}, 25 (2009); I. B\"{u}sching, I., O.C. de Jager, M.S. Potgieter and C. Venter, Astroph. J. Lett. {\bf 678}, 39 (2008).

\bibitem{som1}
M. Cirelli and A. Strumia, Preprint arXiv:0808.3867.

\bibitem{nonthermal}
P. Grajek, G. Kane, D. J. Phalen, A. Pierce and S. Watson, Preprint arXiv:0807.1508.

\bibitem{serpico}
P.D. Serpico, Phys. Rev. {\bf D79}, 1302 (2009).

\bibitem{bland}
R. Blandford and D. Eichler, Phys. Rep. {\bf 154}, 1 (1987).

\bibitem{bell78}
A.R. Bell, MNRAS {\bf 182}, 147 (1978).

\bibitem{kamae}
T. Kamae {\it et al.}, Astroph. J. {\bf 647}, 692 (2006) [Erratum: Astroph. J. {\bf 662}, 779 (2007)].

\bibitem{maldru}
M.A. Malkov and L. O'C. Drury, Rep. Prog. Phys. {\bf 64}, 429 (2001).

\bibitem{KN}
M. Mori, Preprint arXiv:0903.3260.

\bibitem{atic}
J. Chang {\it et al.}, Nature (London) {\bf 456}, 362 (2008).

\bibitem{hess_ele}
F. Aharonian {\it et al.} [HESS Coll.], Phys. Rev. Lett. {\bf 101}, 261104 (2008).

\bibitem{antip}
P. Blasi and P. Serpico, Preprint arXiv:0904.0871.

\bibitem{fermi}
Fermi/LAT Coll., Preprint arXiv:0905.0025.

\end{thebibliography}
\end{document}